\documentclass[iop, apj]{emulateapj}
\usepackage{natbib}
\usepackage{mathptmx} 
\usepackage[T1]{fontenc}
\usepackage{graphicx}

\def\kms {{\mathrm{km}\,\mathrm{s}^{-1}}}

\defcitealias{Pereira:2012spic}{\mbox{Paper I}}

\def\CaII{\ion{Ca}{2}}

\def\MgII{\ion{Mg}{2}}
\def\MgIIk{\ion{Mg}{2}\,k}

\def\hk{{h\&k}}

\def\MgIIhk{\MgII\, \hk}

\def\kthree{\mbox{k$_3$}}    

\def\ktwo{\mbox{k$_2$}}

\def\kone{\mbox{k$_1$}}     
\def\hone{\mbox{h$_1$}}     
\def\ktwoV{\mbox{k$_{2V}$}}
\def\ktwoR{\mbox{k$_{2R}$}}

\def\bifrost{{\it Bifrost}}

\usepackage{mathptmx} 
\usepackage[colorlinks=true,urlcolor=blue,citecolor=blue,linkcolor=blue]{hyperref}

\defcitealias{Leenaarts:Mg2}{Paper I}
\defcitealias{Pereira:Mg3}{Paper II}

\shortauthors{Pereira et al.}

\begin{document}

\title{The formation of \emph{IRIS} diagnostics. IV. The \MgII\ triplet lines as a new diagnostic for lower chromospheric heating}
  
   \author{Tiago M. D. Pereira$^1$}\email{tiago.pereira@astro.uio.no}
   \author{Mats Carlsson$^1$}  %
   \author{Bart De Pontieu$^{2, 1}$}  %
   \author{Viggo Hansteen$^1$}

\affil{$^1$ Institute of
  Theoretical Astrophysics, University of Oslo, P.O. Box 1029
  Blindern, N--0315 Oslo, Norway}
\affil{$^2$Lockheed Martin Solar and Astrophysics Laboratory, 3251 Hanover Street, Org. A021S, Bldg. 252, Palo Alto, CA 94304, USA}

\date{Received; accepted}

\begin{abstract}

A triplet of subordinate lines of \MgII\ exists in the region around the \hk\ lines. In solar spectra these lines are seen mostly in absorption, but in some cases can become emission lines. The aim of this work is to study the formation of this triplet, and investigate any diagnostic value they can bring. Using 3D radiative magnetohydrodynamic simulations of quiet Sun and flaring flux emergence, we synthesize spectra and investigate how spectral features respond to the underlying atmosphere. We find that emission in the lines is rare and is typically caused by a steep temperature increase in the lower chromosphere (above 1500~K, with electron densities above 10$^{18}$~m$^{-3}$). In both simulations the lines are sensitive to temperature increases taking place at column masses $\gtrsim 5\cdot10^{-4}\;\mathrm{g}\:\mathrm{cm}^{-2}$. Additional information can also be inferred from the peak-to-wing ratio and shape of the line profiles. Using observations from NASA's Interface Region Imaging Spectrograph we find both absorption and emission line profiles with similar shapes to the synthetic spectra, which suggests that these lines represent a useful diagnostic that complements the \MgIIhk\ lines.

\end{abstract}

  
\section{Introduction}                          \label{sec:introduction}

The \MgIIhk\ resonance lines are among the strongest in the solar spectrum. They are formed higher than the widely-studied \CaII~H \& K lines owing to the higher magnesium abundance. However, because they sit in the UV spectrum they are not observable from the ground and have not been routinely observed in the past. To observe them astronomers have used a multitude of space platforms, from balloons to space missions \citep[e.g.][]{Durand:1949, Bates:1969, Staath:1995, Doschek:1977, Morrill:2008, West:2011}. The advent of the IRIS mission \citep{IRIS-paper} has provided unprecedented continuous time series of \MgII\ spectra (and slit-jaw filtergrams) with high spatial, spectral, and temporal resolution, which no observatory before could do concurrently. This wealth of \MgII\ spectra is making a difference in how the chromosphere is understood and has great potential for the future.

To understand the complex formation of the \hk\ lines in the Sun, several studies have been undertaken, starting with the early work of \citet{Dumont:1967}, \citet{Milkey:1974}, \citet{Ayres:1976}, \citet{Gouttebroze:1977}, and \citet{Uitenbroek:1997}. More recent work has focused on understanding their formation over a range of solar positions \citep{Avrett:2013}, their polarization potential \citep{Belluzzi:2012}, their formation in prominences \citep{Heinzel:2014}, and their diagnostic value using 3D radiative magneto-hydrodynamic (rMHD) models \citep{Leenaarts:Mg1, Leenaarts:Mg2, Pereira:Mg3}. All of these studies focused on the \hk\ lines; notably absent was any detailed investigation on the companion triplet of lines between the $3p\:^2\!P$ and $3d\:^2\!D$ states \citep[an exception is][who studied these lines above the solar limb]{Feldman:1977}, with vacuum wavelengths of 279.160, 279.875, and 279.882 nm (henceforth referred to as triplet lines).

 With their lower levels being the upper levels of the \hk\ lines (see Figure 1 of \citealt{Leenaarts:Mg1}), the triplet lines sit on the wings of the \hk\ lines (one line on the blue wing of the k line, and two overlapping lines located between the k and the h line). This transition structure bears some resemblance to that of the \CaII\ atom, whose infrared triplet lines (849.8, 854.2, and 866.2 nm) share the upper level of the H\&K lines. With the \CaII\ infrared triplet firmly established as a chromospheric diagnostic in recent literature (in particular the 854.2~nm line, see e.g. \citealt{Cauzzi:2008}, \citealt{Reardon:2009}, \citealt{Leenaarts:2009aa}, \citealt{de-la-Cruz-Rodriguez:2013} and references therein), a study of the diagnostic potential of the \MgII\ triplet is both timely and relevant.

The \MgII\ triplet lines are generally much weaker than their famous siblings \hk\, and appear mostly as absorption lines. Nevertheless, in energetic events they become emission lines. Given the gap in the literature on these lines, in this work we set forth to answer the following questions: in which conditions do these triplet lines form, and what can we learn from them? Under what circumstances do they become emission lines? The lines have already been extensively observed by IRIS (the lines around 279.88~nm are included in virtually all IRIS observations), so any insight can be useful for a wide range of observations. To understand the formation of the lines we employ 3D rMHD models, in a similar way to what \citet{Leenaarts:Mg2} and \citet{Pereira:Mg3} did for the \hk\ lines.

The outline of this paper is as follows. In \S\ref{sec:spectra}, we describe the simulations used and how the synthetic spectra were calculated, and in \S\ref{sec:results} we study how the \MgII\ triplet lines are typically formed in quiet Sun. In \S\ref{sec:emission} we investigate the conditions that lead to emission in the triplet lines, both in a quiet Sun simulation and a flaring simulation. In \S\ref{sec:obs} we show some examples of emission profiles observed with IRIS, and we conclude with a discussion in \S\ref{sec:conclusions}.

\section{Synthetic spectra}                          \label{sec:spectra}

To study the formation of the \MgII\ triplet, we follow the approach of \citet{Leenaarts:Mg2} and \citet{Pereira:Mg3} and make use of 3D rMHD simulations performed with the \emph{Bifrost} code \citep{Gudiksen:2011}. 

\bifrost\ solves the resistive MHD equations on a staggered Cartesian grid. A $24\times 24\times16.8$ Mm$^{3}$ region of the solar atmosphere was simulated, with a constant horizontal cell spacing of 48~km and non-uniform vertical spacing, extending from 2.4~Mm below the average $\tau_{500}=1$ height and up to 14.4~Mm, covering the upper convection zone, photosphere, chromosphere, and lower corona. The radial curvature of the Sun is neglected. The simulations employed include a detailed radiative transfer treatment including coherent scattering \citep{Hayek:2010}, a recipe for non-local thermodynamical equilibrium (non-LTE) radiative losses from the upper chromosphere to the corona \citep{Carlsson:2012}, and thermal conduction along magnetic field lines \citep{Gudiksen:2011}. We use two simulation snapshots from different runs: an ``enhanced-network'' quiet Sun simulation and an emerging flux simulation with some small flares. The quiet Sun simulation snapshot is the same as used in other papers of this series \citep[see][and references therein for more details]{Leenaarts:Mg2}, which includes non-equilibrium hydrogen ionization in the equation of state \citep{Leenaarts:2007}. The photospheric mean unsigned magnetic field strength of the simulation is about 5~mT (50~G), concentrated in two clusters of opposite polarity, placed diagonally 8~Mm apart in the horizontal plane.

We also use a snapshot from the ``flaring'' simulation of \cite{Archontis:2014}. In this emerging flux simulation a uniform magnetic flux sheet with $B_y=336$~mT and of dimension $24\times 13$Mm$^2$ is injected into a numerical domain of $24\times 24\times 17$~Mm that contains a weak initial field of $B<10^{-5}$~mT. The model has fully developed convection and a certain percentage of the injected field emerges into the chromosphere and corona in a non uniform manner, leading to patchy reconnection, as loops expanding through the photosphere into the upper atmosphere come into contact with each other. The reconnection leads to structures resembling small flares. We use the snapshot at $t=10\,000$~s in which several such flares are present. Further details of this simulation setup can be found in \cite{Archontis:2014}.

The synthetic spectra were calculated using the RH1.5D code \citep{RH15D}, a modification of the RH code \citep{Uitenbroek:2001} that solves the non-LTE problem for each column in a 3D atmosphere as an independent 1D column. As shown by \citet{Leenaarts:Mg1} this is a good approximation for the \MgIIhk\ lines, outside the h$_{3}$ and k$_{3}$ cores. To reduce the computational costs, we performed the calculations for every other spatial point in the horizontal directions. We find that the effects of partial redistribution (PRD) in the triplet lines are negligible, therefore we assumed complete redistribution (CRD) for all calculations of these lines, while calculating the \hk\ lines in PRD \citep[using the fast angle-dependent approximation of][]{Leenaarts:2012}.

\section{Formation of the Mg II triplet in quiet Sun} \label{sec:results}

\begin{figure*}
\begin{center}
\includegraphics[width=0.495\textwidth]{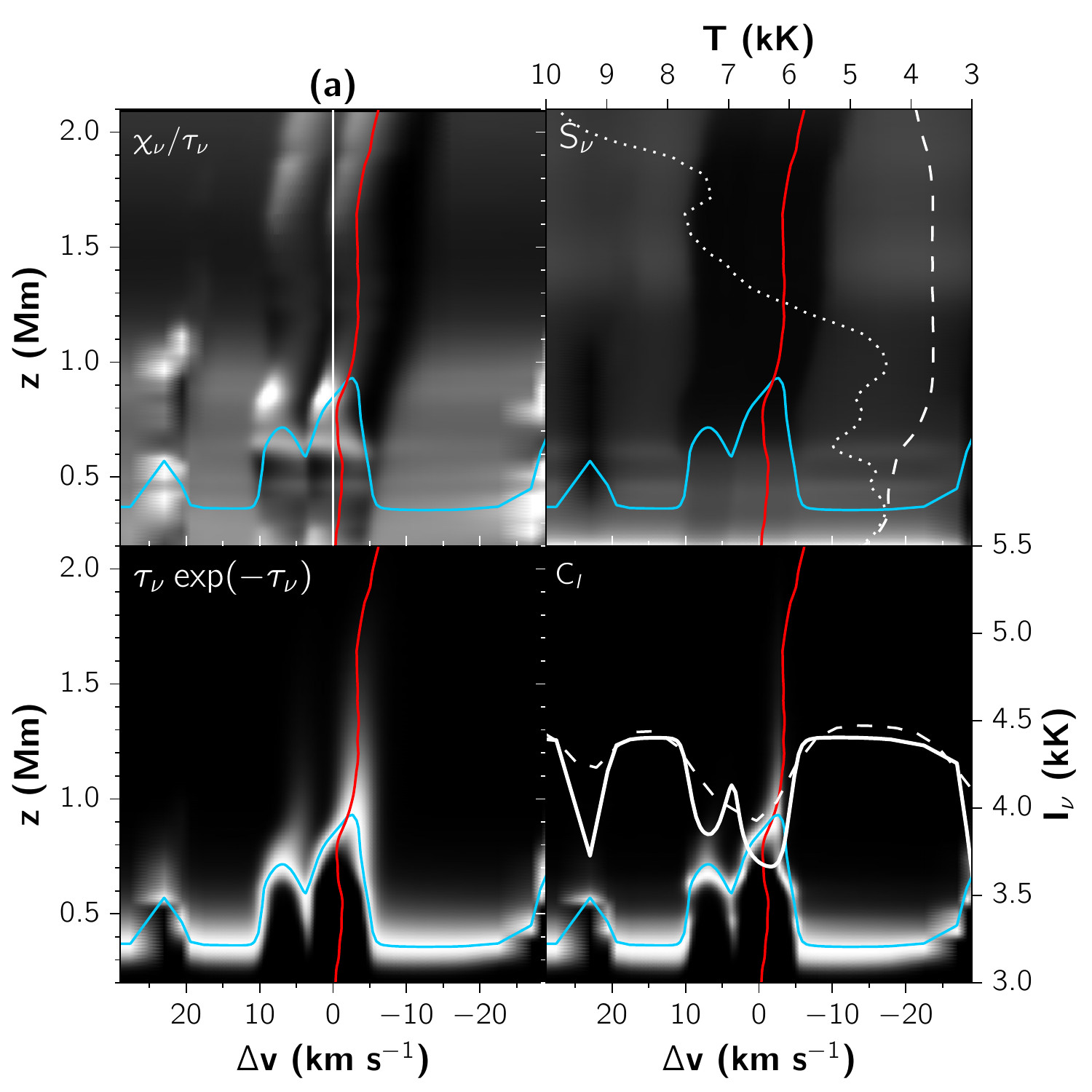}
\includegraphics[width=0.495\textwidth]{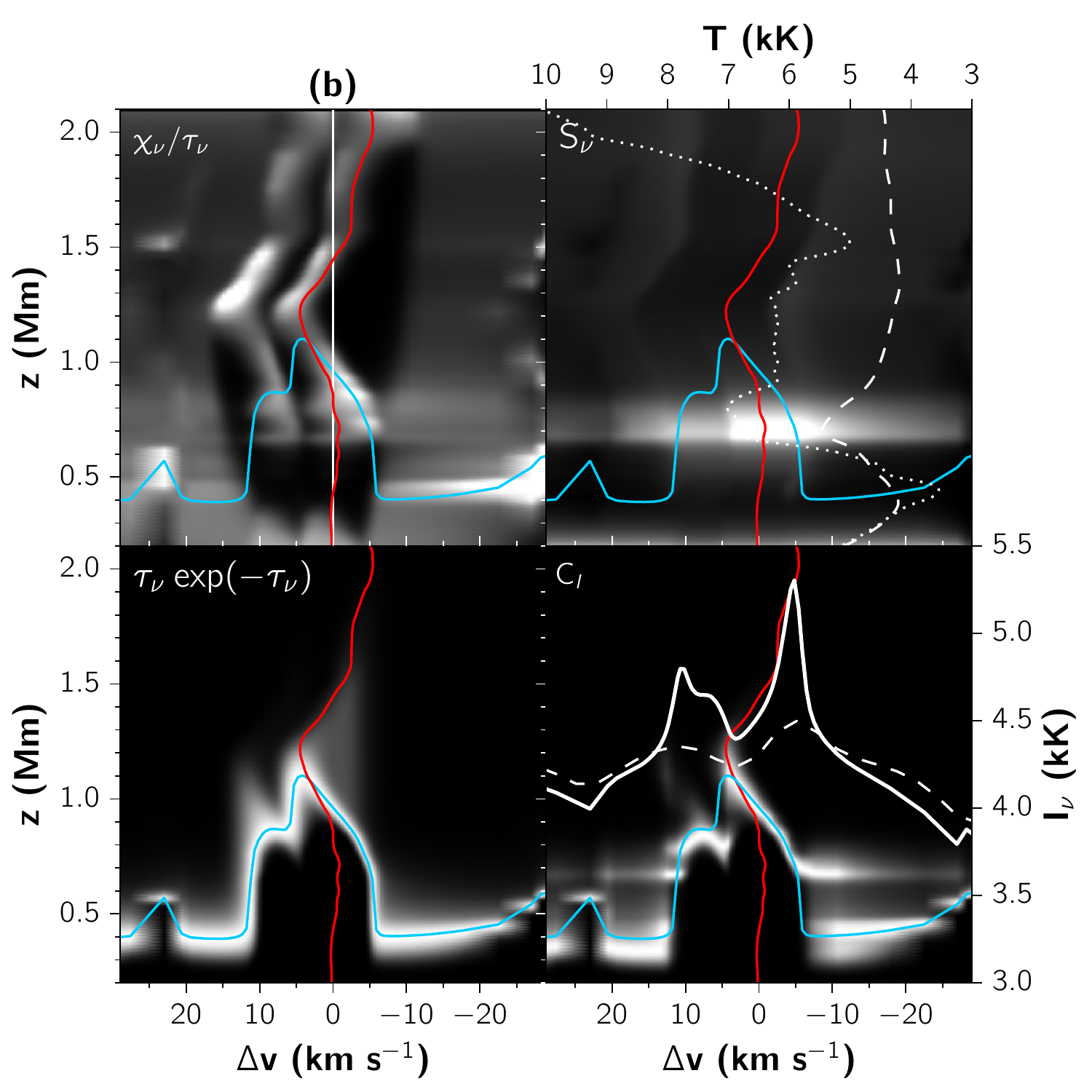}
\end{center}
\caption{Intensity formation diagram for the two \MgII\ lines around 279.88~nm, in two columns of the quiet Sun simulation. Case (a) shows a typical column (left side), with the lines in absorption, while case (b) shows a column where the lines are in emission (right side). For (a) and (b), the individual panels show the quantity specified in the top left corner, where $\chi_\nu$ is the absorption coefficient, $\tau_\nu$ the optical depth, and $S_\nu$ the total source function. The quantities are shown as a function of Doppler shift from 279.88~nm and simulation height $z$. The top left panels ($\chi_{\nu}/\tau_{\nu}$) represent the asymmetry contribution. The bottom right panel shows $C_I$, the contribution function to intensity, and is obtained by multiplying the other three panels (to improve its visibility, $C_I$ was divided by the maximum at each wavelength). A $\tau=1$ curve (cyan) and the vertical velocity (red, positive is upflow) are plotted in each panel, with a $v_z=0$ line in the first panel for reference. In the upper right panel we show also the Planck function (white dotted) and the source function at $\Delta v=0$ (white dashed) in temperature units (scale at the top). The lower right panel contains also the intensity profile in brightness temperature units (scale on the right), at the resolution of the simulation (solid white line) and convolved with the instrumental profile of IRIS \citep[dashed white line, spatially and spectrally convolved as in][]{Pereira:Mg3}. \label{fig:form_diag_qs}}
\end{figure*}

\begin{figure}
\begin{center}
\includegraphics[scale=0.9]{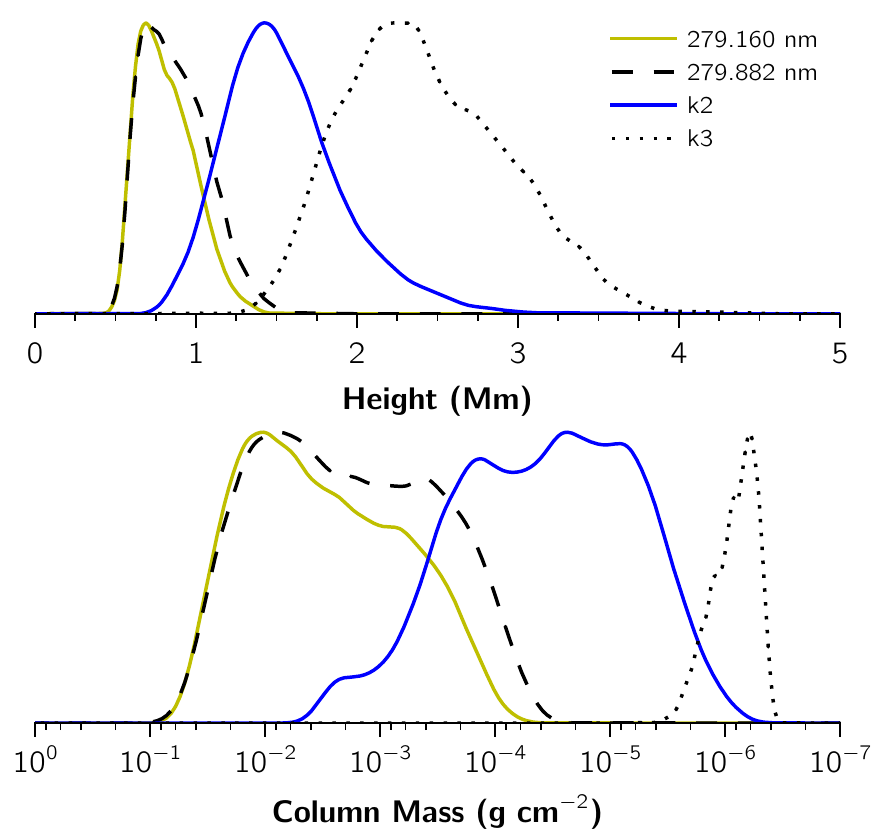}
\end{center}
\caption{Comparison of quiet Sun formation regions for the \MgII\ triplet lines and the k line. Showing distributions of the height where optical depth reaches unity and corresponding column mass, for the line center wavelengths of two triplet lines, and for the \kthree\ and \ktwo\ features of the k line. Curves depict the kernel density estimation with a Gaussian kernel (see text), normalized by the maximum value.\label{fig:histograms_k}}
\end{figure}

In the quiet Sun, the \MgII\ triplet lines are observed mostly as absorption lines \citep[see e.g.][]{Morrill:2008}. This is also the case in our synthetic spectra from the quiet Sun simulation: the lines are in emission in $\approx0.5$\% of the points. The lines are formed at around $0.6-1.2$~Mm above the height where $\tau_{500}=1$. The 279.160~nm line is formed at slightly lower heights than the other pair, which are blended and are indistinguishable at the spectral resolution of IRIS -- they usually appear as a single, wide absorption feature. Compared to the 279.160~nm line, the 279.882~nm line has an oscillator strength twice as high; in terms of formation height this translates to a difference of less than 100~km in most cases. The 279.875~nm line is the weakest, which makes the blend in this region asymmetric, with a centroid shifted toward the red. 

The lines exist in a heavily blended region. We synthesized the strongest nearby lines (assuming LTE) and find that the contribution of lines other than \MgII\ is negligible throughout the line profiles, except in the far wings of the \MgII\ 279.160~nm line.

\begin{figure}
\begin{center}
\includegraphics[scale=0.9]{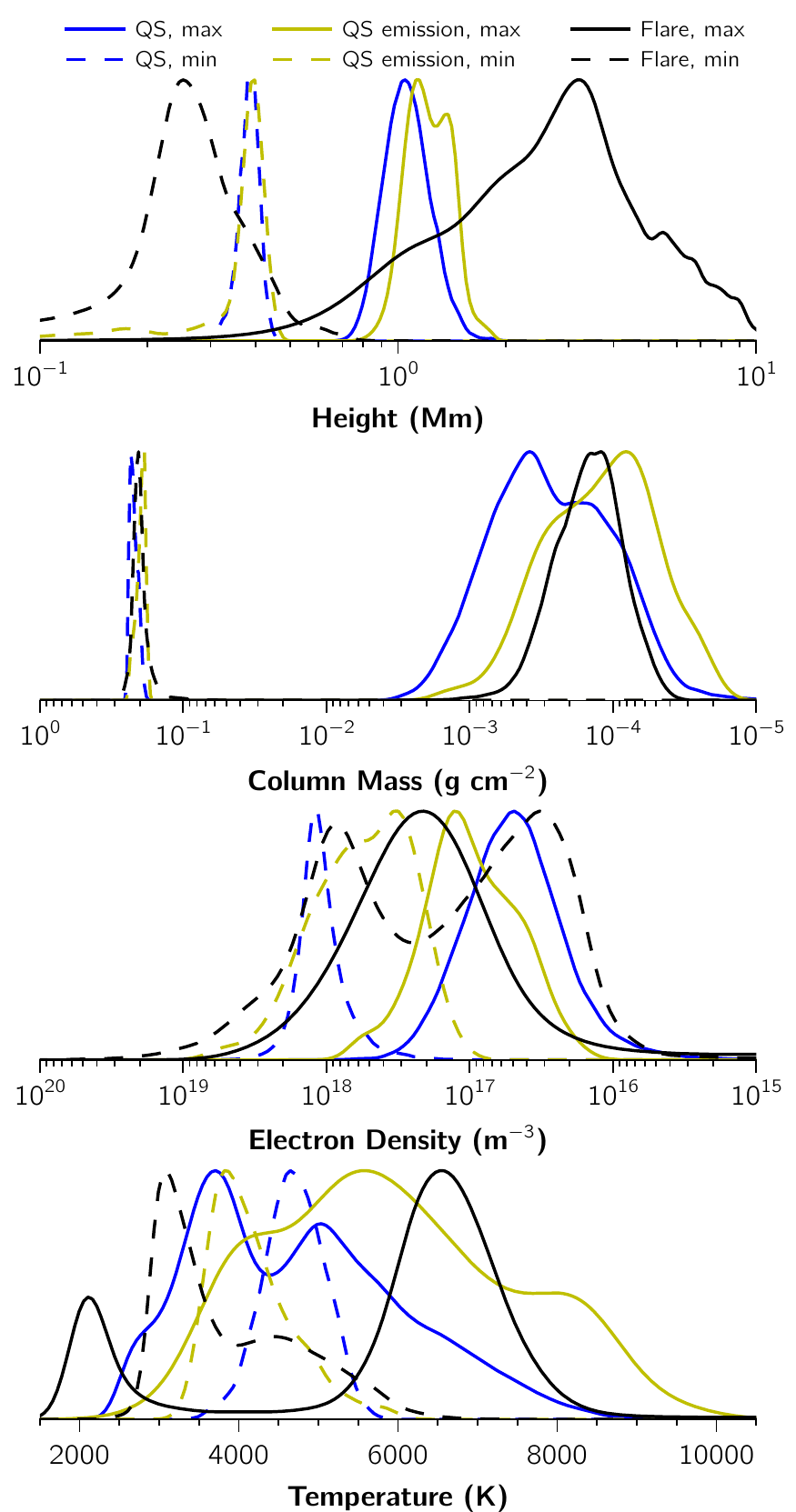}
\end{center}
\caption{Distributions of physical conditions for the maximum and minimum height of unity optical depth, for a spectral range of $\pm 30\;\kms$ around the two strongest \MgII\ triplet lines. Curves depict the kernel density estimation with a Gaussian kernel, normalized by the maximum value. Results are shown for the quiet Sun (QS) simulation for all columns (blue), and for the columns with emission in the triplet lines (yellow), as well as for the flaring simulation (all columns, black).\label{fig:histograms}}
\end{figure}

In quiet Sun conditions, the triplet lines are unremarkable compared to other strong lines in the region \citep[see][]{Pereira:Mg3}. They probe similar atmospheric layers to other strong lines in between the \hk\ lines. The line pair at 279.88~nm is problematic to measure because the lines overlap. Its asymmetric shape complicates the task of measuring velocities or widths with standard techniques (e.g. position of centroid or line fitting). Strong shifts can also make the lines overlap with other nearby lines. In Figure~\ref{fig:form_diag_qs} we analyze the formation of the 279.88~nm lines in two columns from the quiet Sun simulation, in the format developed by \citet{CarlssonStein:1997}. Case (a) shows a typical quiet Sun profile. The two overlapping triplet lines at 279.88~nm are the two humps around $\Delta\mathrm{v}=0$, while the feature at $\Delta\mathrm{v}\approx 23\;\kms$ is a \ion{Fe}{1} line. Here it can be seen that the source function loosely follows the Planck function up to a height of about 0.3~Mm, and then continues to drop, causing an absorption line. (For this location, the \hk\ lines decouple from the source function at $z\approx0.8$~Mm, and still follow the lower part of the temperature increase until $z\approx1.5$~Mm.) Case (b) shows one of the rare cases of emission in the quiet Sun simulation. Here the source function dips with the temperature minimum around $z\approx0.3$~Mm, but then follows a temperature increase and peaks at around $z\approx0.7$~Mm, causing an emission line.

In Figure~\ref{fig:histograms_k} we compare the formation region of the two strongest triplet lines with that of the \MgII~\kthree\ and \ktwo\ features \citep[calculated in the same manner as in][]{Leenaarts:Mg2} for the quiet Sun simulation.  The distributions are computed using Kernel Density Estimation (KDE, see \citealt{rosenblatt1956}, \citealt{parzen1962}) using a Gaussian kernel. In the case of \ktwo\ we took the average $z(\tau=1)$ and column mass of the \ktwoV\ and \ktwoR\ features. One can see how the \MgIIk\ features and the triplet lines cover distinct regions in the range of $0.5-3$~Mm, or $10^{-1}-3\cdot10^{-7}\;\mathrm{g}\:\mathrm{cm}^{-2}$ and therefore can complement each other as diagnostics of the chromosphere.

\section{Emission in the Mg II triplet}             \label{sec:emission}

\subsection{Conditions for emission}

Under particular conditions the triplet lines become emission lines. This seems to happen when there is a rapid increase of temperature with height in the region around the temperature minimum. With such temperature rises, the source function is still close to the Planck function and follows its rise before dropping down in higher layers, causing an emission line. In the synthetic spectra, emission in the line cores is very rare in the quiet Sun simulation ($\approx0.5\%$ of the columns) but much more common in the flaring simulation ($\approx90\%$ of the columns). The conditions for emission become clear in Figure~\ref{fig:form_diag_qs}, where in panel (b) a steep temperature increase at $0.3<z<0.8$~Mm leads the source function to peak and then drop, causing an emission line.

Here we define ``emission'' as the intensity in the line core being higher than in the line wings. In the quiet Sun simulation one finds about 1\% of the columns with emission ``bumps'' in the far wings -- these correspond to locations where the heating takes place at deeper layers and have essentially the same formation mechanism as the profiles with emission in the line core. Nevertheless, for the remainder of this section we restrict ourselves to the more extreme events of line core emission.

In Figure~\ref{fig:histograms} we show distributions for the ranges of heights, column mass, and electron densities where the lines are formed. Unlike in Figure~\ref{fig:histograms_k}, where we show the distributions for $z(\tau=1)$, here we show the ranges of typical conditions where the lines are formed. The distributions are given for the maximum and minimum $z(\tau=1)$, the heights where the optical depth reaches unity, and for the values of column mass, electron density and temperature at those maximum and minimum heights. The bulk of the line is formed in the region between the maximum and minimum distributions, with the line wings formed close to the minimum heights, and the line center formed closer to the maximum heights. Reflecting the very different density profiles, the lines are formed much higher in the emerging flux flaring simulation than in the quiet Sun simulation. However, when shown on a column mass scale the formation region is very similar. The electron density and temperature distributions are more convoluted, and again reflect the differences between the simulations. In the flaring simulation there is a subset of columns with very cool mid-chromospheres, and this causes a double peaked distribution for the temperature of maximum height of formation. There is a clear tendency for the regions in emission to be formed at lower column mass densities, higher up in the atmosphere; this is true for both the quiet Sun simulation and the flaring simulation. We find that the triplet lines are sensitive to column masses of $5\cdot10^{-4}\;\mathrm{g}\:\mathrm{cm}^{-2}$ and higher, in regions with electron densities between $10^{16}-10^{19}\;\mathrm{m}^{-3}$.

\begin{figure}
\begin{center}
\includegraphics[scale=0.9]{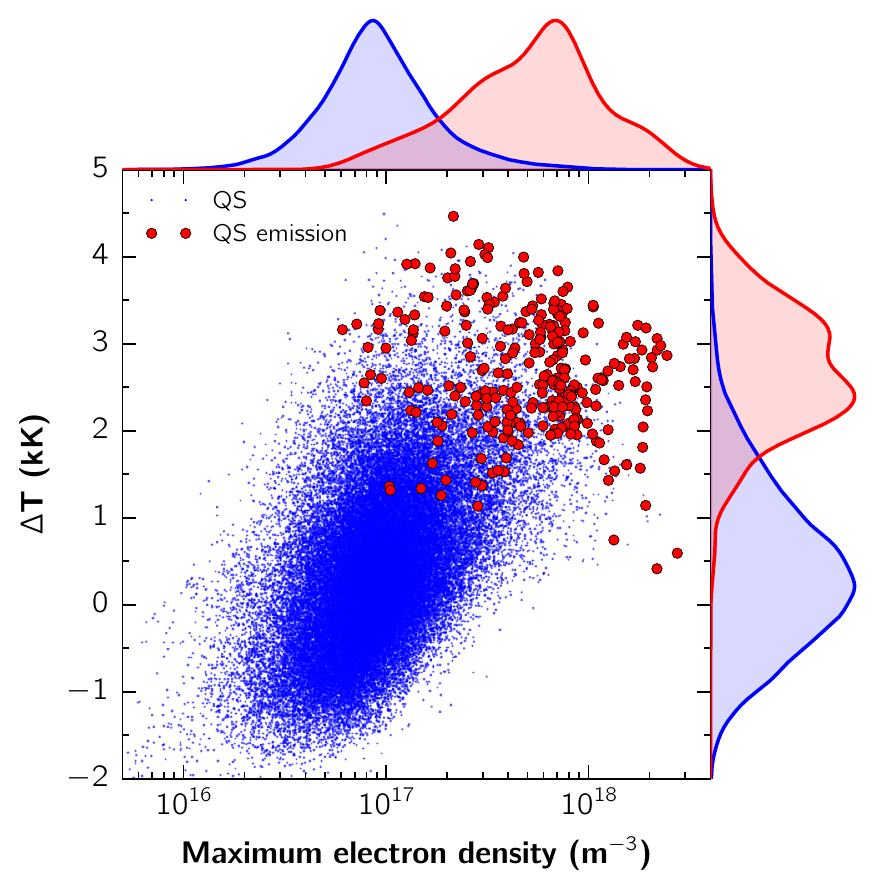}
\end{center}
\caption{Physical conditions for emission in the strongest \MgII\ triplet lines, from the quiet Sun simulation. Temperature difference $\Delta T$ in the line formation region (see text) is plotted against the maximum electron density in the region where line core is formed. Showing all simulation locations (small blue dots) and locations showing emission in the lines (red circles). The top and right diagrams show the distributions (Gaussian kernel density estimation, normalized) for the electron density and $\Delta T$, respectively. \label{fig:tdiff_ne}}
\end{figure}

\begin{figure}
\begin{center}
\includegraphics[scale=0.9]{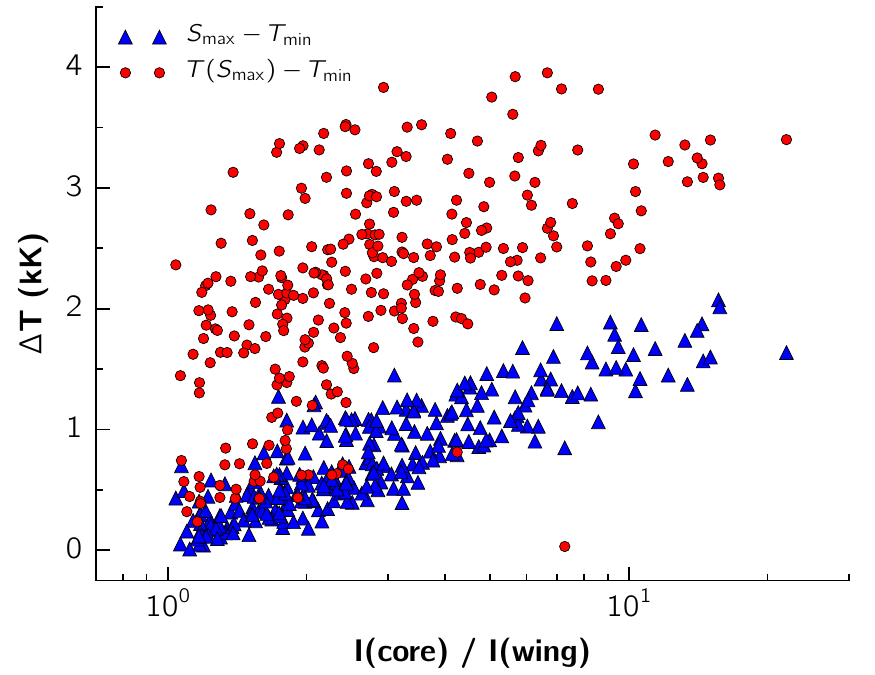}
\end{center}
\caption{Relation between the ratio of line core to wing intensity (for the strongest of the \MgII\ triplet lines) vs.\ $\Delta T$, a temperature difference corresponding to $S_\mathrm{max}-T_{\mathrm{min}}$ (blue triangles) or $T(S_\mathrm{max})-T_{\mathrm{min}}$ (red circles), where $T_{\mathrm{min}}$ is the temperature minimum between column masses of 10 to 0.1~g~cm$^{-2}$, $S_\mathrm{max}$ is the maximum of the source function (in brightness temperature units) between column masses of 0.5 to 10$^{-3}$~g~cm$^{-2}$, and $T(S_\mathrm{max})$ is the temperature at the height where the source function maximum occurs (all points for locations in emission). \label{fig:tdiff_s}}
\end{figure}

\begin{figure*}
\begin{center}
\includegraphics[width=0.495\textwidth]{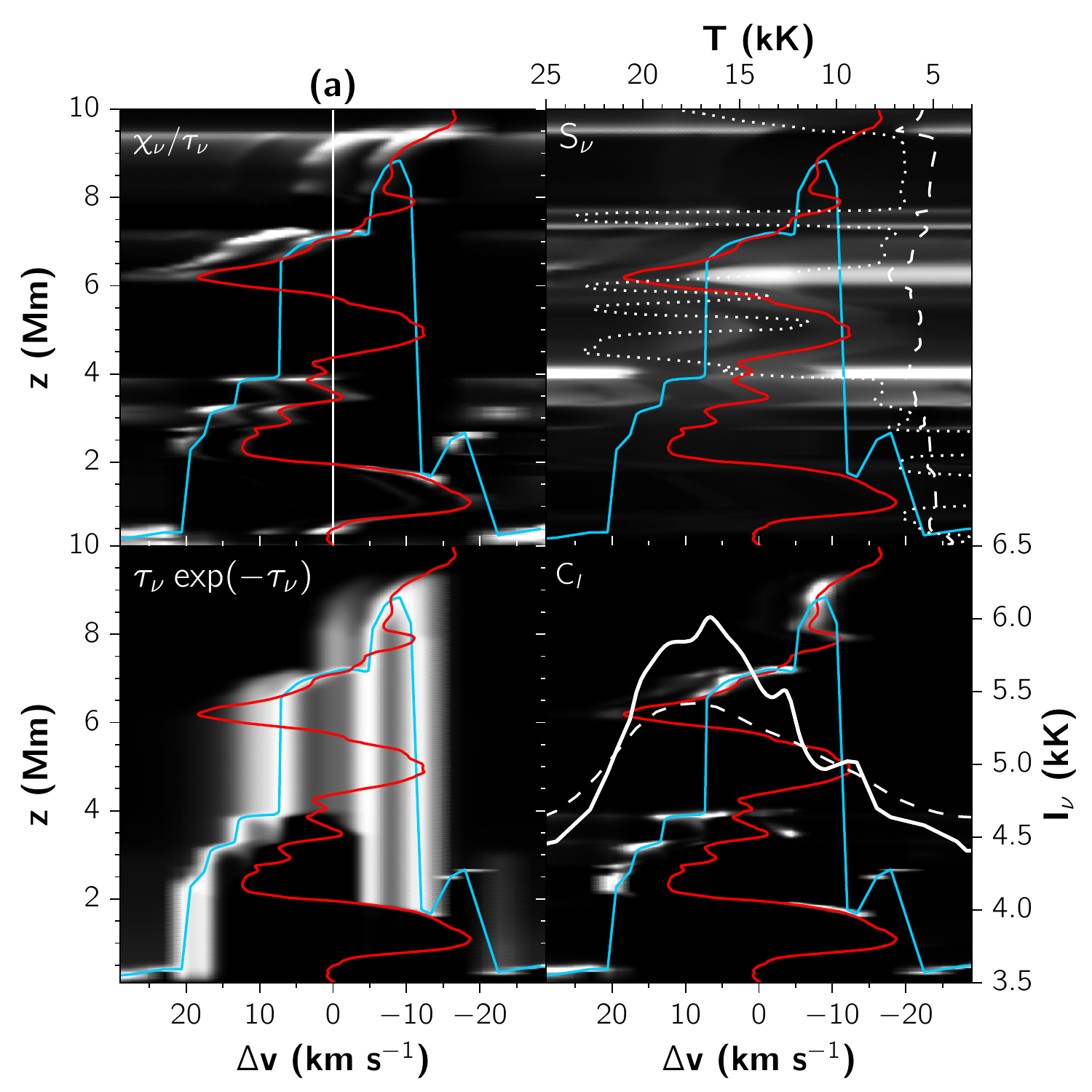}
\includegraphics[width=0.495\textwidth]{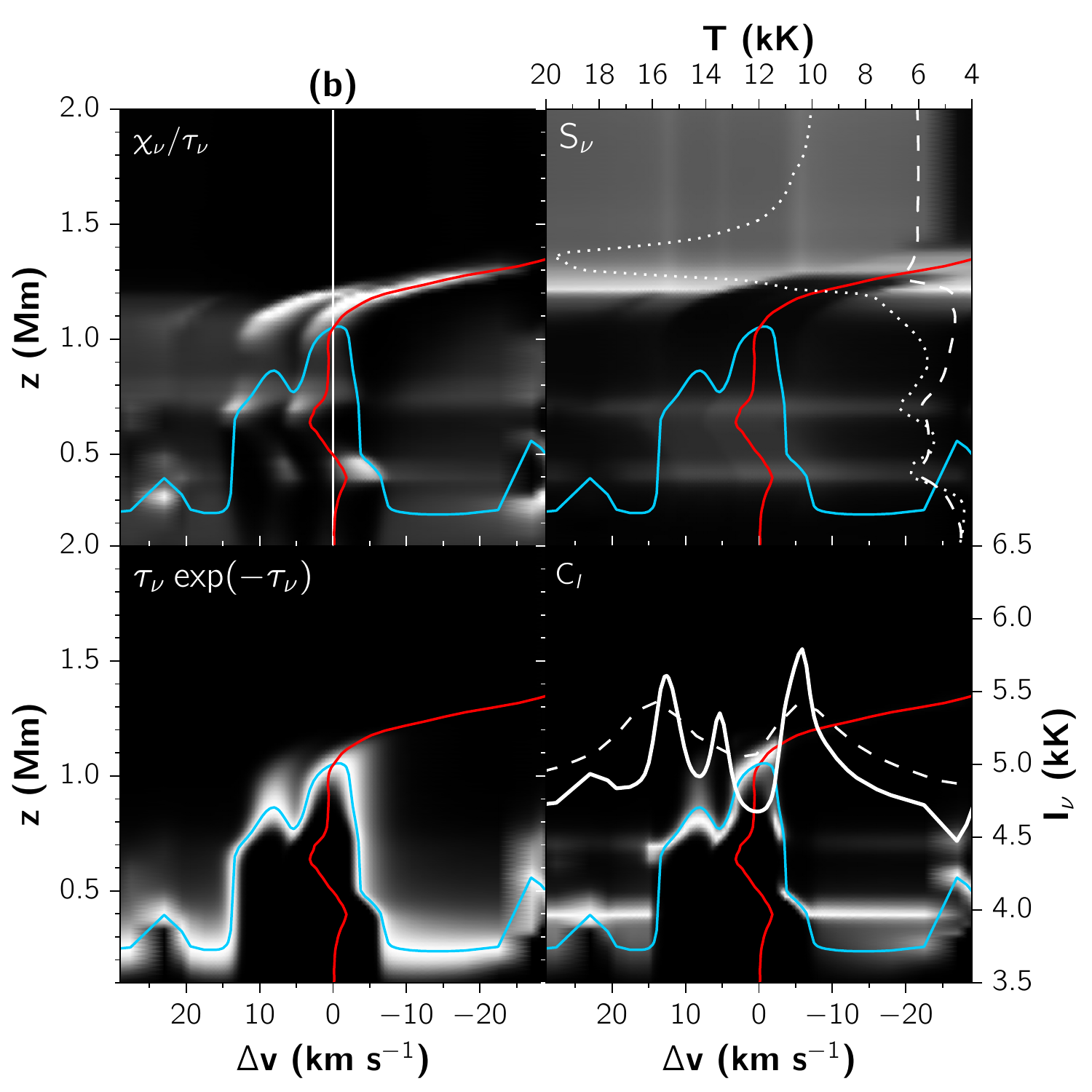}
\end{center}
\caption{Intensity formation diagram for the two \MgII\ lines around 279.88~nm. Cases (a) and (b) correspond to two different columns from the flaring simulation. The format description is the same as for Figure~\ref{fig:form_diag_qs}. Note the different height and temperature scales between (a) and (b).\label{fig:form_diag_flare}}
\end{figure*}

The emission in the lines is caused by a source function increase in the formation region of the line centers. Typically, this increase in the source function is caused by a large temperature increase in the lower chromosphere ($\gtrsim 1500$~K). Nevertheless, there are many cases when large increases do occur, but the source function does not follow the temperature and the lines are in absorption. While there are many factors that affect the coupling between the source function and the local temperatures, the source function tends to be more tightly coupled to the temperature in regions with higher electron density. In Figure~\ref{fig:tdiff_ne} we show a scatter plot and distributions of $\Delta T$ versus the maximum electron density, for the quiet Sun simulation. For each simulation column, $\Delta T$ is the difference between the maximum temperature between the column masses of $5\cdot10^{-4}$ and $10^{-2}\;\mathrm{g}\:\mathrm{cm}^{-2}$ and the minimum temperature between the column masses of $10^{-1}$ and $10\;\mathrm{g}\:\mathrm{cm}^{-2}$ (in other words, the temperature difference between the line core forming regions and the line wing forming regions). The maximum electron density $n_e$ is taken between the column masses of $5\cdot10^{-4}$ to $10^{-2}\;\mathrm{g}\:\mathrm{cm}^{-2}$ (the line core forming region). The columns with emission lines appear clearly clustered in regions with a large temperature difference and high electron density (the distribution of $n_e$ for emission columns peaks at around $7\cdot10^{17}$~m$^{-3}$).

\subsection{Emission as a quantitative diagnostic}

The presence of emission in the triplet lines can be an important indicator that the lower chromosphere has been heated. In addition, it can also be used to quantify the temperature increase that leads to emission. 

We find that the Eddington-Barbier approximation holds for the triplet lines. Under LTE conditions one would expect the intensity to follow the temperature variations. However, despite some coupling to the local temperature (see discussion above), the source function does not completely follow the quick temperature increases that give rise to emission. Still, we find that one can nevertheless derive a quantitative estimate of the temperature increase from the line intensity, and demonstrate it in Figure~\ref{fig:tdiff_s}. We plot an observable from the two blended triplet lines, the $I_{\mathrm{core}}/I_{\mathrm{wing}}$ intensity ratio between the line core maximum (in the $279.866-279.893$~nm interval) and the line wing (here taken at 279.932~nm) against $\Delta T$, a temperature difference given by different quantities. 
The different $\Delta T$ are differences between the line core and wing forming regions. $\Delta T_T \equiv T(S_\mathrm{max}) - T_\mathrm{min}$ gives the physical temperature difference in the line forming region, while $\Delta T_S \equiv S_\mathrm{max} - T_\mathrm{min}$ is a proxy for the temperature difference as ``measured'' by the source function; the discrepancy between the two is a measure of how the source function departs from the local temperature.

As seen in Figure~\ref{fig:tdiff_s}, both $\Delta T$ are correlated with the observable $I_{\mathrm{core}}/I_{\mathrm{wing}}$. When plotted against $\Delta T_S$, the relation with $\log(I_{\mathrm{core}}/I_{\mathrm{wing}})$ is nearly linear -- a consequence of the Eddington-Barbier approximation: the source function and intensity are closely correlated. When plotted against $\Delta T_S$, one can still mostly recover the linear relation with $\log(I_{\mathrm{core}}/I_{\mathrm{wing}})$, but there is an added offset and increased scatter. For clarity we show only the results for the quiet Sun simulation in Figure~\ref{fig:tdiff_s}, but the results for the flaring simulation are essentially the same. Within the uncertainties, this enables one to use $I_{\mathrm{core}}/I_{\mathrm{wing}}$ to directly quantify the localized heating in the lower chromosphere when the triplet lines are in emission, making it a powerful diagnostic. 

\begin{figure*}
\begin{center}
\includegraphics[scale=0.9]{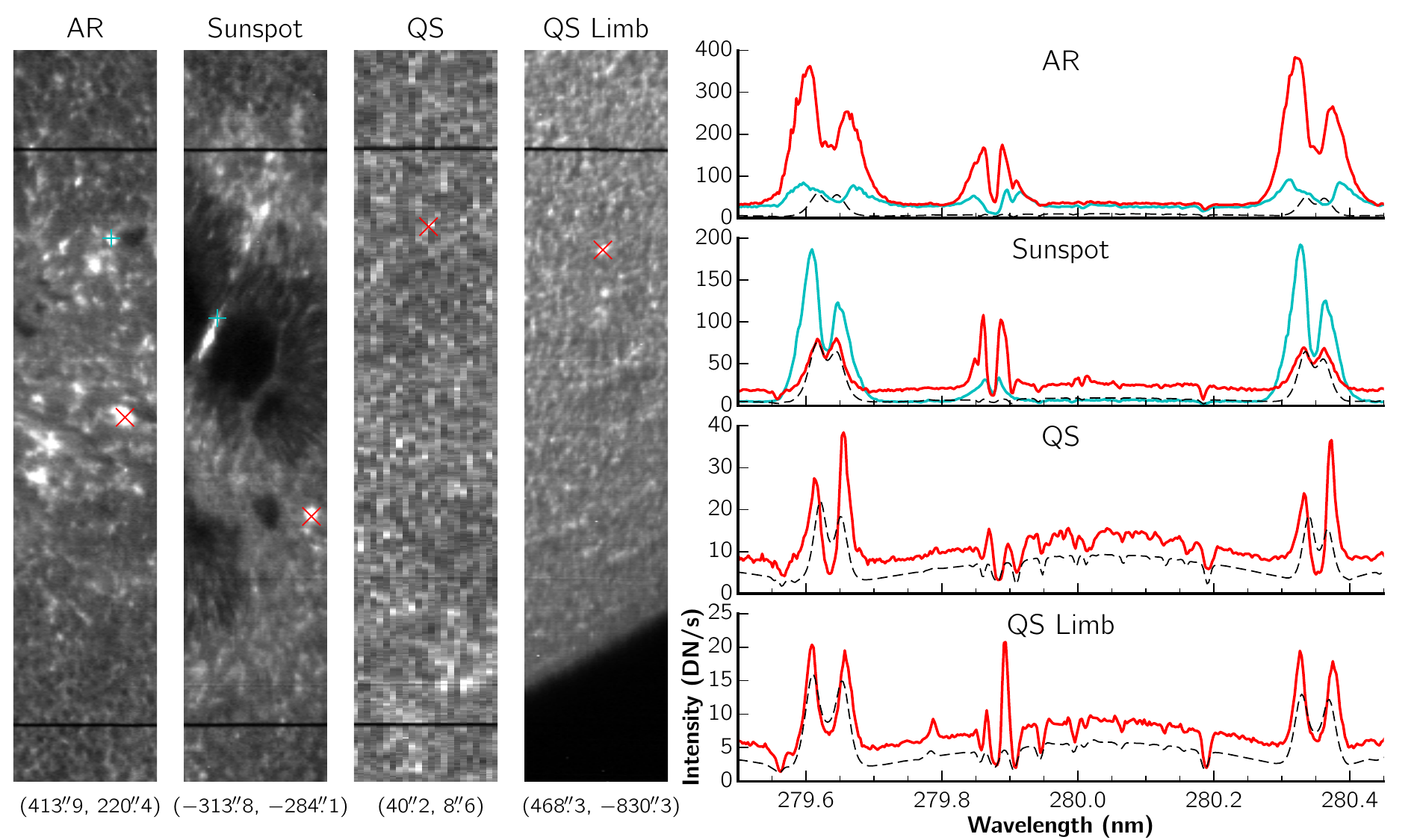}
\end{center}
\caption{Examples of triplet emission in IRIS observations. \emph{Left panels:} images from spectral rasters, taken at fixed wavelengths in the 279.88~nm lines (for each panel, the wavelengths were taken where the peak emission occurs at the spatial position of the red crosses). Rasters were taken in different regions and observing modes (see text for details), including active region (AR), near a sunspot, in the quiet Sun (QS), and in quiet Sun near the limb (QS Limb). Each image is oriented in solar (x, y) coordinates, and has a field-of-view of $22\farcs4\times120\arcsec$; the coordinates of the bottom left corners are shown below the images. \emph{Right panels:} Individual spectra from selected points in the rasters on the left, labelled accordingly. Red curves correspond to the red crosses on the images, cyan curves to the cyan plus signs (only for AR and Sunspot). The black dashed curves depict the spatially averaged spectra over the windows displayed on the left.\label{fig:obs}}
\end{figure*}

In Figure~\ref{fig:form_diag_flare} we show example line formation diagrams for the flaring simulation. Case (a) is an extreme example with strong velocity and temperature gradients alternating in rapid succession in height. The resulting triplet lines are formed over a very large height range and show a strong emission profile. There are two large peaks of the source function at around $z=4, \:6$~Mm, which roughly follow large temperature increases. Near the line core the $z=6$~Mm peak is dominant, making the line profile nearly single peaked. Case (b) depicts a very different scenario: here the line is formed over a shallow height range ($z\lesssim1.3$~Mm), where the velocities are relatively small. In this case there are two temperature bumps at $z\approx 0.4,\:0.7$~Mm, where the densities are high enough that the source function follows. This causes a triple-peaked profile. There is another, much larger, temperature increase at $z\approx 1.3$~Mm and corresponding bump in the source function. However, at these higher layers there is very little opacity at these wavelengths, and correspondingly their contribution to the line profile is negligible. (At  $z= 1.33$~Mm the column mass in this location is $6\cdot10^{-5}\;\mathrm{g}\:\mathrm{cm}^{-2}$, just outside the range where the lines are sensitive.)

As evidenced by the line profiles in Figures~\ref{fig:form_diag_qs} and \ref{fig:form_diag_flare}, the shape of the emission lines also provides important information about the underlying physical conditions. Double, triple, or multi-peaked profiles are mostly caused by a temperature profile with several rapid variations in height. Single-peaked lines tend to come from a dominant temperature increase, typically in higher layers. If a single temperature increase occurs deep in the atmosphere, at high densities, the emission manifests itself in the far wings of the triplet lines, with the line core a normal absorption profile (when this happens, the \hk\ lines are also much wider than normal). In the quiet Sun simulation, there are in fact twice as many columns with far wing emission profiles than columns with emission in the line core.

\section{Observing emission with IRIS}                   \label{sec:obs}

The \MgII\ triplet lines are routinely observed with the IRIS spectrograph. In Figure~\ref{fig:obs} we show some IRIS raster images and individual spectra of locations with emission in the triplet lines. The observations comprise active regions (AR, one of them including a sunspot) and quiet Sun (both at disk-center and at the limb). Details about the observations are summarized in Table~\ref{tab:obs}. The observations were obtained by scanning a region of the Sun with a moving slit, with a step size of $0\farcs35$ (``dense'' rasters), except the quiet Sun dataset, which has a step size of $1\arcsec$. We made use of IRIS reduced and calibrated level 2 data \citep[see][for details on the reduction procedure]{IRIS-paper}.

\begin{deluxetable}{lllr}
\tablecaption{Observational data sets.\label{tab:obs}}
\tablehead{
\colhead{Target} & \colhead{Starting time} & \colhead{Center coord.} & \colhead{Exp. time}}
\startdata
Active Region & 2014-06-15T08:29 & ($419\farcs8$, $280\farcs4$) & $4$~s \\
Sunspot & 2014-07-03T10:10 & ($-302\farcs6$, $-225\farcs0$) & $4$~s \\
Quiet Sun & 2013-09-18T07:39 & ($34\farcs9$, $68\farcs0$) & $4$~s \\
Quiet Sun limb & 2014-07-10T00:10  &  ($538\farcs3$, $-770\farcs4$) & $30$~s 
\enddata
\end{deluxetable}

From the observations we find that emission in the triplet lines is rare. They are most easily seen in emission in flares and other high-energy phenomena where there is heating in the lower chromosphere. \citet{Vissers:2015} find that the triplet lines are often in emission in Ellerman bombs \citep{Ellerman:1917}, and they are also seen in emission in some of the explosive events reported by \citet{Peter:2014} and \citet{Schmit:2014}.

The raster images in Figure~\ref{fig:obs} are taken at wavelengths close to the center of the lines at 279.88~nm, meaning that bright areas are locations of increased intensity in the lines (in the AR and sunspot images many of these are indeed in emission, but not in the quiet Sun images). In the AR images one sees a collection of several bright dots with a round shape where the lines are in emission -- these generally occur in the vicinity of sunspots, and it is possible that they are related to Ellerman bombs. But as shown in the sunspot panel, there is also strong triplet emission (and strong \hk\ emission) in the light bridge. In some of these AR locations the \hk\ lines have an intensity very close to the average, while the triplet lines are strongly enhanced (see the extreme example in the sunspot spectrum where the lines at 279.8~nm are stronger than the k line). Such scenarios could indicate an abrupt temperature rise only in the lower chromosphere, not felt by the \hk\ line centers. Other locations show both the triplet lines and the \hk\ lines strongly enhanced, which could be caused by a temperature increase throughout a wider range of the chromosphere (e.g. in flares). 

In the quiet Sun, emission in the triplet lines is seldom found. From several datasets investigated, it was found only in very few of them. And even when found, it is rarely of the same magnitude as seen in ARs. Most often, this quiet Sun emission is on the far wings of the line, with the central part of the line resembling its typical absorption profile. This is consistent with what we found in the quiet Sun simulation, and suggests that in these locations the heating is limited to a deep area near the temperature minimum. Another piece of corroborating evidence for this scenario is that the \hk\ lines are also noticeably wider than the average, an indication that the \kone/\hone features are being formed lower down in the atmosphere as a result of the chromospheric temperature increase taking place deeper than usual. This is the case with the quiet Sun profiles that we show -- the triplet lines are in emission only in the wings, and the \hk\ are wider than normal. These locations also show an enhanced photospheric temperature, as evidenced by the high local continuum.

\section{Conclusions}                            \label{sec:conclusions}
We have studied the formation of the \MgII\ triplet of lines that lie near the \hk\ lines. The lower levels of these subordinate lines are the upper levels of the \hk\ lines. To understand their formation we use a forward modeling approach, making use of realistic 3D radiative MHD simulations and comparing the predicted spectra with the physical quantities from the simulation. 

In the quiet Sun, we find that the lines are formed just above the temperature minimum, at heights around $0.6-1.2$~Mm above $\tau_{500}=1$, or at column masses down to $10^{-4}\;\mathrm{g}\:\mathrm{cm}^{-2}$. The lines can be used as velocity indicators for those layers, but the line pair at 279.88~nm is problematic because of the overlapping lines. The source functions of these lines typically decouple from the local temperature before the chromospheric temperature rise, and therefore the source function decreasing with height gives rise to absorption lines, by far the most common scenario. However, under particular circumstances the lines go into emission, and this can be a powerful diagnostic.

When a large temperature increase is present in the lower chromosphere the source functions can follow this increase, leading to emission lines. From the simulations we find that this typically happens when there is a temperature increase of more than 1500~K in layers with column masses from 1 to $10^{-3}\;\mathrm{g}\:\mathrm{cm}^{-2}$, and an electron density above $10^{17}\;\mathrm{m}^{-3}$. In addition, one can use the ratio of the emission peak to the local continuum of the lines to derive a rough estimation of the temperature difference that caused that same emission. This holds true for both quiet Sun and more violent flaring simulations, and means that the lines can be used to diagnose steep temperature increases in the lower chromosphere, a new type of diagnostic complementary to those of the \hk\ lines \citep{Leenaarts:Mg1, Leenaarts:Mg2, Pereira:Mg3}. The shape of the emission line also provides information about the underlying physical quantities: in cases where the heating occurs deeper down and is covered by cooler material, the emission in the triplet lines occurs predominantly in their far wings, with the central part of the line being like a typical absorption line (under these circumstances the \hk\ lines are also wider). When the heating occurs higher in the column range to which the lines are sensitive ($10^{-2}-5\cdot10^{-4}\;\mathrm{g}\:\mathrm{cm}^{-2}$), the emission takes place in the line core.

The \MgII\ triplet lines are routinely observed by IRIS, and we find several example observations that confirm the scenarios seen in the synthetic spectra. As in the quiet Sun simulation, emission is very rare in the quiet Sun, and when it happens it tends to be in the far wings of the lines, suggesting heating occurring deeper in the chromosphere. The lines are more easily observed in emission in flares, active regions, and in particular near sunspots and features like Ellerman bombs, as has already been reported. In such locations the lines can be strongly enhanced, in some extreme cases even stronger than the \hk\ lines. With the help from the simulations, one can now understand better some of these phenomena and use the diagnostics from these lines to trace instances of strong heating in the lower chromosphere. 

\acknowledgments{
   We would like to thank Masoumeh Izadparast for her help in finding triplet emission 
   in the IRIS observations, and the referee for many useful comments.
   This research was supported by the
   Research Council of Norway through the grant ``Solar Atmospheric
   Modelling'' and by grants of computing time from the Programme for Supercomputing,
   and by the European Research Council under the European 
   Union's Seventh Framework Programme (FP7/2007-2013) / ERC Grant 
   agreement No. 291058.
   This work has benefited from discussions at 
   the International Space Science Institute (ISSI) meeting on 
   ``Heating of the magnetized chromosphere'' from 5-8 January, 2015, 
   where many aspects of this paper were discussed with other colleagues.
   B.D.P. acknowledges support from NASA grants NNX08AH45G, 
   NNX08BA99G, NNX11AN98G, NNM07AA01C (\emph{Hinode}), and NNG09FA40C (\emph{IRIS}). 
   IRIS is a NASA Small Explorer mission developed and operated by LMSAL with 
   mission operations executed at NASA ARC and major contributions to downlink
   communications funded by the NSC (Norway).
}

\bibliographystyle{apj}

\end{document}